\documentstyle[12pt,amsmath]{article}
\begin{document}
\begin{titlepage}
\begin{flushright}
HD-THEP-2003-05
\end{flushright}
\vspace{1.8cm}
\begin{center}
{\bf\LARGE Cosmology with varying scales and couplings}\\
\vspace{1cm}
C. Wetterich\footnote{e-mail: C.Wetterich@thphys.uni-heidelberg.de}\\
\bigskip
Institut  f\"ur Theoretische Physik\\
Universit\"at Heidelberg\\
Philosophenweg 16, D-69120 Heidelberg

\vspace{1cm}
\today
\end{center}
\begin{abstract}
The time variation of fundamental mass scales can have profound cosmological implications.
We investigate a particular model of crossover quintessence which is compatible with all
present cosmological observations. This model can also reconcile the reported time
variation of the fine structure constant from quasar absorption lines with
the bounds from archeo-nuclear physics and tests of the equivalence principle.
\end{abstract}

\end{titlepage}
\begin{center}
{\bf Dynamical mass scales}
\end{center}

The recent confirmation of dark energy by the high precision KMAP data \cite{KMAP}
challenges for an answer to the question if dark energy is constituted by a
static cosmological constant or by dynamical quintessence \cite{CW1}. A dynamical
dark energy may shed light on the profound question of the role and origin of mass
scales in physics. It may also lead to a cosmological time variation of dimensionless
fundamental parameters like the electromagnetic fine structure constant.

In quantum field theory or statistical physics the dimensionless couplings depend on length
scales \cite{GML}. This effect is perhaps most striking in QCD where the running gauge
coupling generates the mass scale for the hadrons. The modern view of the renormalization
group \cite{KW} associates this running with an ``integrating out'' of fluctuations as one
looks at the effective physics on larger and larger length scales. The fluctuation effects
can be treated in a conceptual simple and elegant form in the context of an exact functional
renormalization group equation for the effective average action \cite{ERGE}. It describes
for both renormalizable and nonrenormalizable theories the dependence of the fluctuation
effects on an effective infrared cutoff. The physical length scale which cuts off the
fluctuations in the infrared may correspond to the expectation value of a scalar field
\cite{CWg} and therefore become dynamical. We explore here the hypothesis that the time
evolution of scales persists until the present epoch in the universe \cite{CW1}. This would
have striking effects for cosmology, the time variation of fundamental ``constants'' and
tests of the equivalence principle.

For the renormalization of the standard model of particle physics one usually assumes
implicitely the existence of an ultraviolet scale - typically the Planck scale or a
grand unification scale. Similarly, for solids the ultraviolet scale is given by the
lattice distance or for gases and liquids by the molecular size. In contrast, we explore
here the concept of a unified theory where the action does not involve any explicit
fundamental mass scale \footnote{In string theories the apparent mass scale corresponding
to the string tension can be replaced by the expectation value of the dilaton, thereby
making this scale dynamical as well.}. Even without knowing the fundamental unified theory
the experience with quantum physics lets us expect two types of possible mass scales in such
a scenario. First, some intrinsic mass scales may be generated by the running of
dimensionless couplings, similar to the hadron masses in QCD. Second, the expectation
values of scalar fields induce dynamical mass scales. This second type of scales can
change in the course of the cosmological evolution.

We explore here the scenario that the Planck mass or Newtons constant is dynamical. In
contrast to the early formulation of this idea by Dirac and Jordan \cite{D}, \cite{J} we
postulate \cite{CW2}, \cite{CW1} that the particle masses are dynamical as well,
such that in a first
approximation the ratio between the nucleon mass and the Planck mass, $m_n/\bar{M}_p$,
remains fixed, and similar for the weak scale $M_W/\bar{M}_p$, the electron mass
$m_e/\bar{M}_p$ etc. In this first approximation {\em all} mass scales are proportional
to a scalar field $\chi$, i.e. $\bar{M}_p=\chi,m_n=h_n\chi$ or $m_e=h_e\chi$. This changes
profoundly both the conceptual status and the observational consequences as compared
to scenarios with fixed particle masses \cite{J,BD}.

In fact, in absence of intrinsic mass scales generated by the running dimensionless
couplings the effective action (after integrating out all quantum fluctuations) would
exhibit an exact dilatation symmetry, corresponding to a multiplicative rescaling
$\chi\rightarrow\alpha\chi$. If we neglect in a first approximation the intrinsic
mass scales, any nonzero value of the scalar field $\chi$ would break the exact global
dilatation symmetry. In consequence of the spontaneous breaking of scale symmetry
most particles would acquire a mass $\sim\chi$. Another consequence would be a Goldstone
boson $\sim\ln\chi$ which has only derivative interactions - the dilaton. In this
approximation of an exact spontaneously broken dilatation symmetry no interesting
consequences  for late cosmology are expected \cite{CW2,CW1}. However, the impact of
the dilaton on cosmology changes dramatically if the effect of the possible intrinsic
mass scales is taken into account. In terms of the dilatation symmetry the intrinsic mass
scales reflect an ``anomaly'' induced by the quantum fluctuations. The scalar field $\chi$
can now constitute a dynamical dark energy or quintessence \cite{CW1} which is relevant
for the present universe. Because of its important cosmological consequences the
``pseudo-dilaton'' $\ln\chi$ has been named the cosmon.

We present here a particular version of crossover quintessence \cite{CQ} with two
intrinsic mass scales generated by running dimensionless couplings. The first mass scale
$\chi_c$ indicates a crossover behavior in the derivative interactions of the pseudo
Goldstone boson, without introducing a potential. More concretely, it affects the kinetic
term of the cosmon and indirectly the location of an ultraviolet fixed point for the running
grand unified gauge coupling. The second scale $m$ is analogous to the hadron scale
$\Lambda_{QCD}$ in QCD. It is an infrared scale where the running grand unified gauge
coupling grows large. This scale $m$ induces a mass term for the cosmon field
$\sim m^2\chi^2$. Similar to the ratio $\Lambda_{QCD}/\bar{M}_p$ the ratio
$m/\chi_c$ can be exponentially small. We will see that for $\ln(m/\chi_c)\approx 1/138$
a realistic cosmology arises, with quintessence responsible for about $70\%$ of the
present energy density of the universe. Our model turns out to be
compatible with all present cosmological tests.

Furthermore, this model predicts that the fundamental ``constants'' depend on time
\cite{CW2,CW1}. In fact, the running of a typical dimensionless coupling like the
fine structure constant $\alpha_{em}$ can be written in the functional form
$\alpha_{em}(q^2/\chi^2,\chi^2/\mu^2)$ where $\mu$ is an arbitrary renormalization scale.
For fixed $\chi$ the dependence on the squared momentum $q^2$ is governed by the usual
$\beta$-function of the standard model. On the other hand, for fixed $q^2/\chi^2$
the variation of $\chi$ is
governed by the running of the grand unified gauge coupling $\alpha_X$. The cosmological
time dependence of $\chi$ results in a time dependence of the fine structure costant!
We show that for reasonable $\beta$-functions for the running grand unified gauge coupling
our model can reconcile the claimed time variation from the observation of quasar
absorption lines \cite{Webb} with severe bounds from archeo-nuclear physics
\cite{Oklo,Re} and tests of the equivalence principle \cite{EP}.

\newpage
\begin{center}
\textbf{Crossover Quintessence}
\end{center}

Crossover quintessence (CQ) can be characterized by a recent increase of the fraction in
homogeneous dark energy $\Omega_h=\rho_h/\rho_{cr}$ from a small but not negligible value
for early cosmology to a value $\Omega_h\approx0.7$ today. A small early value (say
$\Omega_h=0.01)$ typically results from a cosmic attractor solution \cite{CW1,Rat,CW3,Att}
(``tracker solution'') independently of the detailed initial conditions. For such models
of ``early quintessence'' the dark energy decreases in early cosmology at the same pace as
radiation. Dark energy and radiation have therefore always been of a comparable
magnitude and excessive ``fine tuning'' can be avoided \cite{CW1,Heb}. The amount of early
quintessence is accessible to observation by nucleosynthesis \cite{CW1}\cite{NS}, the cosmic
microwave background (CMB) \cite{Doran} or structure formation \cite{FJ,Schwindt}. The
recent crossover to a domination of the universe by dark energy ($\Omega_h\approx0.7$) may
be related to properties of the underlying model \footnote{This typically requires a modest
``tuning'' of parameters at the level somewhat below 1\%.} or triggered by matter domination
\cite{Muk} or structure formation \cite{BR}. On the level of the equation of state CQ
corresponds to a crossover from a value $w_h\approx1/3$ during the radiation dominated
universe to a substantially negative value for $z<0.5$.

In this note we elaborate on a recent proposal \cite{CQ} that CQ is connected to the
existence of a conformal fixed point in a fundamental theory. This will serve as a natural
starting point for a discussion of the variation of couplings. We assume that for low momenta
$(q^2\ll\chi^2)$ the yet unknown unified theory results in an effective
grand unified model in four dimensions.
There the coupling of the cosmon field $\chi$ to gravity and the gauge fields is described
by an effective action
\begin{eqnarray}\label{3}
S&=&\int d^4x\sqrt{g}\left\{-\frac{1}{2}\chi^2R+\frac{1}{2}(\delta(\chi)-6)\partial^{\mu}
\chi\partial_{\mu}\chi\right.\nonumber\\
&&\left.+m^2\chi^2+\frac{Z_F(\chi)}{4}F^{\mu\nu}F_{\mu\nu}\right\}.
\end{eqnarray}
The running of the effective grand unified gauge coupling $g_X^2=\bar{g}^{2}/Z_F$ is determined
by the $\chi$-dependence of $Z_F$ and we assume a fixed point behavior for
$\alpha_X=g^2_X/4\pi$
\begin{equation}\label{4}
\frac{\partial\alpha_X}{\partial\ln\chi}=b_2\alpha_X-b_4\alpha^2_X-b_6\alpha^3_XS(\delta).
\end{equation}
We take here a small value of $|b_6|$ such that the last term only leads to a small shift
in the fixed point value
\begin{equation}\label{5}
\alpha_{X,*}\approx\frac{b_2}{b_4}\approx\frac{1}{40}.
\end{equation}
(We choose positive $b_2$ and $b_4$ and $S(\delta)$ will be specified below.) For
$\alpha_X(\chi)>\alpha_{X,*}$ the grand unified gauge coupling increases with decreasing
$\chi$. Similar to QCD it grows large at some nonperturbative scale (``confinement scale'')
that we associate with $m$. Our model therefore has an ``infrared scale'' $m$ which is
generated by dimensional transmutation from the running of the gauge coupling. We assume
that $m$ determines the mass of the cosmon. Neglecting the effects of the dilatation anomaly
reflected by $m$ the cosmon potential would vanish.

For $\chi\gg m$ the gauge coupling runs towards its fixed point (\ref{5}) as $\chi$ increases.
At the fixed point the effective action would have an effective dilatation symmetry if the
running of $\delta(\chi)$ could be neglected. We will assume, however, that the running of
$\delta$ is instable towards large $\chi~(E>0)$ and postulate
\begin{equation}\label{6}
\frac{\partial\delta}{\partial\ln\chi}=\beta_\delta(\delta)=E\delta^2.
\end{equation}
For small enough $\delta_i=\delta(\chi=m)$ the dimensionless coupling $\delta$ stays small
for a large range $m<\chi\le\chi_c$. For $\chi\gg m$ we may neglect the small mass $m$ for the
evolution of the dimensionless couplings. The evolution in this range is then governed by the
vicinity to a conformal fixed point \cite{CQ} at
$\delta=\delta_*=0~,~\alpha_X=\alpha_{X,*}$.
In the same approximation our model corresponds to a flat direction \footnote{In string
theories such flat directions are often called ``moduli''.  In presence of additional matter
fields, e. g. scalar and fermions, the flat direction in the space of scalar fields
corresponds to constant ratios of all masses \cite{CQ}.} in the effective potential,
consistent with conformal symmetry. All cosmon interactions are therefore given by its
derivative coupling to the graviton and gauge fields. We note that the fixed point behavior
bears a certain resemblance with the ``runaway dilaton''. \cite{Dam}

Nevertheless, as $\chi$ increases further an effective ``ultraviolet scale'' $\chi_c$
characterizes the region where $\delta$ grows large according to the solution
\begin{equation}\label{7}
\delta(\chi)=\frac{1}{E\ln(\chi_c/\chi)}.
\end{equation}
The ratio between the ultraviolet and infrared scales turns out exponentially huge
\begin{equation}\label{8}
\frac{\chi_c}{m}=\exp\left(\frac{1}{E\delta_i}\right).
\end{equation}
At the crossover scale $\chi_c$ the flow of the couplings witnesses a crossover from
the vicinity of the conformal fixed point (small $\delta$) to another regime (perhaps
a different fixed point) for large $\delta$. We will see that the crossover behavior
in the present epoch of the cosmological evolution precisely corresponds to the crossover
from small to large $\delta$ for $\chi$ in the vicinity of $\chi_c$. Realistic cosmology
obtains for $E\delta_i\approx 1/138$.

The precise behavior of $\beta_{\delta}$ in the region of large $\delta$ will only be of
secondary relevance for past and present cosmology provided $\delta$ grows large enough
near $\chi_c$. For the present note we investigate first a scenario where the increase of
$\delta$ with $\chi$ is unbounded while $\delta(\chi)$ is defined for all $\chi$, namely
\begin{equation}\label{9}
\beta_{\delta}=\frac{E\delta^2}{1+J_{\delta}\delta}.
\end{equation}
(We take $J_{\delta}=0.05$ and observe that cosmology shows no strong dependence on the
value of $J_{\delta}$. This holds provided that $J_{\delta}$ is small enough such that
the difference between eqs. (\ref{6}) and (\ref{9}) matters only in the region of large
$\delta$. At the end of this note we will compare this scenario with an alternative where
$\beta_{\delta}$ exhibits a fixed point for large $\delta$.) For very large $\delta$ in the
region $\chi\gg\chi_c$ the solution obeys now $\delta(\chi)/\delta(\chi_c)\sim
(\chi/\chi_c)^{E/J_{\delta}}$. We observe that for large $\delta$ a standard scalar kinetic
term obtains for a rescaled scalar field
$\tilde{\chi}\sim\chi\delta^{1/2}(\chi)\sim\chi^{1+E/(2J_{\delta})}$. In this
language the coupling to gravity vanishes for $\tilde{\chi}\rightarrow\infty$ according
to $\chi^2R\sim\delta^{-1}\tilde{\chi}^2R$. Our first scenario for the behavior at large
$\delta$ can therefore also be viewed as a fixed point in the cosmon-gravity coupling
$\delta^{-1}$ at $\delta^{-1}_*=0$.

In the sense of a renormalization group running our first
scenario describes a crossover between two fixed points at $1/\delta=0$ and $\delta=0$.
The scale $\chi$ corresponds to a dynamical infrared scale of the unknown fundamental theory.
As it is lowered the flow of the couplings switches from the fixed point at $\delta^{-1}_*=0$
to the range of attraction of the conformal fixed point at $\delta_*=0$. However, this
latter fixed point is not stable with respect to the running of the gauge coupling which
finally grows large and produces a new infrared scale $m$. This non-perturbative scale
lifts the degeneracy in the flat cosmon direction. This type of behavior is common in
statistical physics when the crossover between two fixed points plays a role.

\begin{center}
{\bf Dark energy}
\end{center}

The cosmological dynamics of CQ is described in \cite{Heb,CWTV} and best expressed after
Weyl scaling with a constant reduced Planck mass $\bar{M}^2=M^2_p/8\pi$ and a rescaled
dimensionless scalar field $\Phi=\varphi/\bar{M}=2\ln(\chi/m)$. In terms of the rescaled
metric and cosmon field the effective action (\ref{1}) reads \cite{Heb}
\begin{equation}\label{10}
S=\int d^4x\sqrt{g}\left\{\frac{\bar{M}^2}{2}\big[k^2(\phi)\partial^{\mu}\phi\partial
_{\mu}\phi-R\big]+\bar{M}^4\exp(-\phi)
+\frac{Z_F(\phi)}{4}F_{\mu\nu}F^{\mu\nu}\right\}
\end{equation}
with \footnote{Eq. (\ref{6}) corresponds to $k^{-2}=2E(\phi_c-\phi)$. Similar to eq.
(\ref{9}) this is a form of ``leaping kinetic term quintessence'' \cite{Heb}.} $k^2(\phi)=
\delta/4$. The cosmon potential $V=\bar{M}^4\exp(-\phi)$ vanishes for $\phi\rightarrow\infty$.
In this language it is easy to see that the asymptotic behavior for increasing time
corresponds to increasing $\phi$ and $\chi$.

The cosmological evolution equations are best
expressed in terms of the logarithm of the scale factor
\begin{equation}\label{11}
x=\ln a=-\ln(1+z)~,~\dot{x}=H.
\end{equation}
The first equation expresses \footnote{We have neglected the effects of the cosmon coupling
to matter and radiation which are assumed to be very small in the present model.} the
change of the energy density of homogeneous quintessence, $\rho_h$, in terms of its
equation of state $w_h=p_h/\rho_h$,
\begin{equation}\label{12}
\dot{\rho}_h+3(1+w_h)H\rho_h=0\hspace{0.5cm},\hspace{0.5cm}
\frac{d\ln\rho_h}{dx}=-3(1+w_h).
\end{equation}
Using $\rho_{cr}=3\bar{M}^2H^2=\rho_h+\rho_m+\rho_r~,~d\ln\rho_m/dx=-3~,~
d\ln\rho_r/dx=-4~,~\rho_r=(a_{eq}/a)\rho_m$, eq. (\ref{12}) can be translated into an
evolution equation for $\Omega_h=\rho_h/\rho_{cr}$, namely
\begin{equation}\label{13}
\frac{d\Omega_h}{dx}=-\Omega_h(1-\Omega_h)
\left\{3w_h-\left((1+\frac{e^x}{a_{eq}}\right)^{-1}\right\}.
\end{equation}
We recover the fixed points for $w_h\approx 1/3$ if $a=e^x\ll a_{eq}$, and $w_h\approx 0$ if
$e^x\gg a_{eq}$. They are relevant for the cosmic attractors in the radiation and matter
dominated epochs.

As a second equation we need the time evolution of $w_h$ for which we find \footnote{
We use $d\phi/dx=k^{-1}\big(6\Omega_h(1-V/\rho_h)\big)^{1/2}$ and $V/\rho_h=(1-w_h)/2$.}
\begin{equation}\label{14}
\frac{dw_h}{dx}=(1-w_h)\left\{k^{-1}(\phi)\sqrt{3\Omega_h(1+w_h)}-3(1+w_h)\right\}.
\end{equation}
The dynamics of the particular model enters through $k^{-1}(\phi)=2/\sqrt{\delta}$. Our
scaling of $\phi$ is chosen such that at all times the value of $\phi$ is directly connected
with the value of the cosmon potential and therefore also with $H^2$,
\begin{equation}\label{15}
\frac{H^2}{\bar{M}^2}=\frac{2}{3}e^{-\phi}[\Omega_h(1-w_h)]^{-1}.
\end{equation}
For given present values of the Hubble parameter $H_0$ and the energy fractions in matter,
radiation, and quintessence, $\Omega^{(0)}_m,~\Omega^{(0)}_r=a_{eq}\Omega^{(0)}_m,~
\Omega^{(0)}_h=1-\Omega^{(0)}_m-\Omega^{(0)}_r$ this expresses $\phi$ in terms of
$x,~\Omega_h$ and $w_h$
\begin{equation}\label{16}
\phi=4x-2\ln(H_0/\bar{M})-\ln\left[\frac{3}{2}(a_{eq}+e^x)\Omega^{(0)}_m\right]
-\ln\frac{(1-w_h)\Omega_h}{1-\Omega_h}.
\end{equation}
The system of the two differential equations (\ref{13})(\ref{14}) is therefore closed and
can easily be solved numerically.

For given $E$ and $J_{\delta}=0.05$ we solve the flow equation (\ref{9}). The ``initial value''
$\delta_i=\delta(\chi=m)$ is tuned such that at present $\Omega^{(0)}_h=0.73$. We start for
$\chi(t=0)=m$ with rather arbitrary values of $\Omega_h$ and $w_h$. The late time behavior
will not depend on this. In table 1 we present
a few characteristic cosmological quantities, namely the present equation of state,
$w^{(0)}_h$, the age of the universe, $t^{(0)}$, the fraction of dark energy at last
scattering, $\Omega^{(ls)}_h$, the location of the third peak in the CMB unisotropies in
angular momentum space, $l_3$, as well as the normalization of the spectrum of density
fluctuations, $\sigma_8$, divided by the value which would obtain for a cosmological
constant with the same $\Omega^{(0)}_h=0.73$. Here the ``present'' values are characterized
by the Hubble parameter reaching its present value with $h=0.71$. We recall that smaller
$h$ and larger $\Omega^{(0)}_h$ shift $l_3$ to larger values.

\vspace{0.5cm}
\noindent
\label{table1}
\begin{tabular}{|c|c|c|c|c|c|}\hline
$E$&
$w^{(0)}_h$&
$t^{(0)}/10^{10}yr$&
$\Omega^{(ls)}_h$&
$l_3$&
$\sigma_8/\sigma^{(\wedge)}_8$\\ \hline
$5$&
$-0.94$&
$1.31$&
$0.019$&
$779$&
$0.70$\\
$12$&
$-0.99$&
$1.35$&
$0.0083$&
$794$&
$0.85$ \\ \hline
\end{tabular}

\vspace{0.5cm}
\noindent
Table 1: Characteristic cosmological quantities for two CQ models.

\vspace{0.5cm}
We have chosen the cosmological parameters $\Omega^{(0)}_h$ and $h$ in accordance with the
best fit of the KMAP-data \cite{KMAP} for the case of a cosmological constant
$(w^{(0)}_h=-1)$. For $E=12$ the values $w^{(0)}_h=-0.99$ and $\Omega^{(ls)}_h=0.008$
imply that for the CMB spectrum there is barely any difference as compared to the case of a
cosmological constant with the same amount of dark energy \cite{Doran}. We emphasize,
however, that the value $\sigma_8/\sigma^{(\Lambda)}_8<1$ indicates that more power on large
scales can be tolerated in order to achieve consistency with the galaxy surveys. This may
make the KMAP data more compatible with a $k$-independent spectral index. We have not tried
to optimize the choice of $\Omega^{(0}_h$ and $h$. For $E=5$ a better agreement with the
KMAP-data may be achieved by varying these parameters.
We conclude that for large enough $E$ our model is consistent with present cosmological
observations \footnote{Adding to $\beta_{\delta}$ a term linear in $\delta$ would further
suppress $\Omega_h$ at early time and bring the phenomenology even closer to the one for a
cosmological constant (for given E). The prize to pay are unnaturally small values of
$\delta_i$, similar to quintessence with an inverse power law potential \cite{Rat}.}.

\begin{center}
{\bf Cosmological phenomenology of crossover quintessence}
\end{center}
For a detailed comparison with cosmological observation one is interested in the
dependence of the Hubble parameter on redshift, $H(z)$, and related quantities. Rather than
testing each particular model - as characterized here by the function $k(\phi)$ - one wants
to describe the function $H(z)$ by a few parameters whose values may be fixed by future
cosmological precision tests. We propose to concentrate for this purpose on the average
equation of state $\big(x=-\ln(1+z)\big)$
\begin{equation}\label{(A)}
\bar{w}_h(x)=-\frac{1}{x}\int^0_xdx^{\prime}w_h(x^{\prime}).
\end{equation}
This function is linked very directly with cosmological observables and admits a simple
parameterization. We believe that for the next generation of observations a quadratic
formula will be sufficient (for $z<10^4$)
\begin{equation}\label{(B)}
\bar{w}_h(x)=w^{(0)}_h+(\bar{w}^{(ls)}_h-w^{(0)}_h)\frac{x}{x_{ls}}+Ax(x-x_{ls}).
\end{equation}
Here the three parameters are the equation of state today, $w^{(0)}_h$, the averaged
equation of state at last scattering, $\bar{w}^{(ls)}_h$ with $x_{ls}=-\ln(1100)$, and $A$.
(In the future, one wants to measure the whole function $\bar{w}_h(x)$ as precisely as
possible - it contains the complete information about the dynamics of dark energy.)

Indeed, from $\bar{w}_h(x)$ one can easily construct the time history of $\Omega_h$
according to
\begin{equation}\label{(C)}
\frac{\Omega_h(x)}{1-\Omega_h(x)}=
\frac{\Omega^{(0)}_h(1+a_{eq})}{1-\Omega^{(0)}_h}
\frac{\exp\big(-3x\bar{w}_h(x)\big)}{1+a_{eq}\exp(-x)}
\end{equation}
with $a_{eq}$ the scale factor at matter-radiation equality. We can also obtain
the equation of state at any time
\begin{equation}\label{(D)}
w_h(x)=\frac{d}{dx}\big(x\bar{w}_h(x)\big).
\end{equation}
The function $H(z)$ is related to $\bar{w}_h(x)$ by
\begin{equation}\label{(E)}
\frac{H^2(z)}{H^2_0}=\Omega^{(0)}_h(1+z)^{3\big(1+\bar{w}_h(x)\big)}
+\Omega^{(0)}_m[(1+z)^3+a_{eq}(1+z)^4]
\end{equation}
where $\Omega^{(0)}_m=(1-\Omega^{(0)}_h)/(1+a_{eq})$. We note that $\bar{w}^{(ls)}_h$ in
eq. (\ref{(B)}) is directly related to the fraction of dark energy at last scattering,
$\Omega^{(ls)}_h$, via eq. (\ref{(C)}). This quantity can be tested by the CMB-unisotropies
\cite{Doran}. For a given model of quintessence we determine A be equating the fit formula
(\ref{(B)}) with the model-computation of $\bar{w}_h$ at $x=x_{ls}/2$. For the models
displayed in table 1 this yields reasonable fits for the range $x_{ls}\leq x\leq 0$, with
$\bar{w}^{(ls)}_h=-0.221,A=-0.016$ for $E=5$ and $\bar{w}^{(ls)}_h=-0.261,A=-0.013$ for
$E=12$. We emphasize that the polynomial fit should not be used too far outside this range
(e.g. for nucleosynthesis) since CQ typically results in a crossover from $\bar{w}_h$
near $1/3$ for large negative $x$ to a constant negative value of $\bar{w}_h$ for large
positive $x$. However, for all observations of recent cosmology from the time of last
scattering until now the proposed fit will do reasonably well. The fact that we choose
$\bar{w}_h$ and a polynomial in $\ln(1+z)$ permits us to cover the whole range between
last scattering and now. Supernovae observations at $z\approx 1$ can therefore be linked to
structure formation and CMB within a simple parameterization. This would be impossible with
a Taylor expansion of $w_h(z)$ in $z$ around $z=0$.

\begin{center}
\textbf{Cosmological history of fundamental couplings}
\end{center}

Under quite general circumstances quintessence
or a time varying dark energy can be associated with a scalar field whose
cosmological expectation value varies during the recent history of the universe \cite{CW1}.
Generically, this scalar field - the cosmon - may also couple to  matter and radiation. As
a consequence, the values of ``fundamental constants'' like the fine structure constant or
the ratio between the nucleon mass and the Planck mass depend on the value of the cosmon field
and therefore on cosmological time \cite{CW2,CW1,CW3,Dvali,Chiba}. Recently, a time variation
of the fine structure constant has been reported \cite{Webb} from the observation of quasar
absorption lines (QSO) at redshift $z\approx 2$. A typical value corresponds to
\begin{equation}\label{1}
\frac{\Delta\alpha_{em}(z=2)}{\alpha_{em}}=-7\cdot10^{-6}.
\end{equation}
Similar observations infer \footnote{A recent collection of results on
$\Delta\alpha_{em}$ as well as the time variation of other fundamental constants can be
found in \cite{Uzan}.} a substantially smaller value of $|\Delta\alpha_{em}|$ at lower
redshift $z\le 0.7$. From the Oklo natural nuclear reactor one
obtains a typical bound \cite{Oklo},
\cite{Re}
\begin{equation}\label{2}
\frac{|\Delta\alpha_{em}(z=0.13)|}{\alpha_{em}}<10^{-7}
\end{equation}
and analysis of the decay rate $Re^{187}\rightarrow Os^{187}$ restricts \cite{Re}
\begin{equation}\label{2a}
\frac{|\Delta\alpha_{em}(z=0.45)|}{\alpha_{em}}<3\cdot10^{-7}.
\end{equation}
As an obvious question one may ask if an increase of $|\Delta\alpha_{em}|$ by almost two
orders of magnitude between $z=0.13$ and $z=2$
can be reasonably explained by quintessence models. In this note we
demonstrate that this can indeed be the case in a class of models of ``crossover
quintessence'' proposed recently \cite{CQ}. This contrasts to constant rates
$\partial\Delta\alpha_{em}/\partial z=const$. or $\partial\Delta\alpha_{em}/\partial t=const.$
for which the values and bounds (\ref{1})-(\ref{2a}) clearly are in discrepancy. The
deviation from constant rates also affects strongly the comparison of eq. (\ref{1}) with
bounds from nuclear synthesis \cite{BBN},\cite{Av} and CBM \cite{Av}.

We are aware that the bound from the Oklo natural reactor is subject to substantial
uncertainties due to possible cancellations between effects from the variation of
$\alpha_{em}$ and other ``fundamental constants'' like the mass ratio between pion
and nucleon mass $m_{\pi}/m_n$ or the weak interaction rates. Also the
QSO observations need further verification and investigation of systematics.
Nevertheless, it is well conceivable that we can
get a reasonable ``measurement'' of the function $\Delta\alpha_{em}(z)$ in a not too
distant future. Furthermore, the derivative $\partial\Delta\alpha_{em}(z)/\partial z$ at
$z=0$ can be related \footnote{See \cite{Dvali}, \cite{Chiba} for a different relation
which gives a much smaller violation of the equivalence principle.}
\cite{CWTV} to precision tests of the equivalence principle \cite{EP}.
The aim of this note is to present a sample computation how the knowledge of
$\Delta\alpha_{em}(z)$ can be used to constrain models of quintessence. For this purpose we
will use the value (\ref{1}) as a ``benchmark value'' which fixes the strength of the cosmon
coupling to matter and radiation. The time history $\Delta\alpha_{em}(z)$ can then be
related to the time history of quintessence.

In particular, we may investigate the ratio
\begin{equation}\label{AA1}
R=\Delta\alpha_{em}(z=0.13)/\Delta\alpha_{em}(z=2)
\end{equation}
A bound $R<1/70$ strongly favors \cite{CWTV}
quintessence with a time varying equation of state $w_h=p_h/\rho_h$, where the value of
$(1+w_h)$ at present is substantially smaller than for $z=2$. A logarithmic dependence of
$\Delta\alpha_{em}$ on the cosmon field $\chi$ leads to a bound
for the present equation of state \cite{CWTV}
\begin{equation}\label{AA2}
w^{(0)}_h<-0.9.
\end{equation}
In this note we investigate
how the detailed dependence of $\alpha_{em}$ on $\chi$ influences the time history of
$\Delta\alpha_{em}(z)$. We argue that reasonable $\beta$-functions for the ``running'' of
a grand unified coupling with $\ln\chi$ may lead to nonlinearities in $\alpha_{em}(\ln\chi)$.
Those are quantitatively important, influencing $R$
within a factor two. The overall picture remains rather solid,
however: a small value of $|R|$ requires a quintessence model where the evolution of the
cosmon field has considerably slowed down in the recent history as compared to a redshift
$z\approx 2$. This feature is characteristic for crossover quintessence and will not be
shared by arbitrary quintessence models. In particular, a constant equation of state with
$w_h$ independent of $z$ over a range $0<z<3$ will have severe difficulties to explain
$R<0.1$. This demonstrates how a measurement of $\Delta\alpha_{em}(z)$ could become an
important ingredient for the determination of the nature of dark energy.

\begin{center}
\textbf{Running couplings}
\end{center}

In a grand unified theory the value of the electromagnetic fine structure constant depends
on the gauge coupling at the unification scale, $\alpha_X$, the ratio between the weak scale
and the unification scale, $M_W/M_X$, and particular particle mass ratios like
$m_n/M_W$ or $m_e/M_W$ for the nucleons and electrons. Since the nucleon mass is determined
by the running of the strong gauge coupling similar dependencies arise for the ratio
$m_n/M_X$. Combining the relevant one loop formula and assuming for simplicity
$\chi$-independent ratios $m_b/M_W$ etc. for the heavy quarks and $m_{\mu}/m_e$ etc. for the
leptons one finds \footnote{We have eliminated the dependence of $\alpha_{em}$ on
$M_X/M_W$ in favor of $M_W/m_n$.} \cite{CWTV}
\begin{eqnarray}\label{17}
\frac{1}{\alpha_{em}(z)}-\frac{1}{\alpha_{em}(0)}&=&\frac{22}{7}
\left(\frac{1}{\alpha_X(z)}-\frac{1}{\alpha_X(0)}\right)\nonumber\\
&&-\frac{17}{21\pi}\ln\frac{(M_W/m_n)(z)}{(M_W/m_n)(0)}+\frac{2}{\pi}\ln
\frac{(M_W/m_e)(z)}{(M_W/m_e)(0)}
\end{eqnarray}
and
\begin{eqnarray}\label{18}
\ln\frac{(m_n/\bar{M})(z)}{(m_n/\bar{M})(0)}&=&-\frac{\pi}{11}\left(\frac{1}{\alpha_{em}(z)}
-\frac{1}{\alpha_{em}(0)}\right)\\
&+&\frac{7}{33}\ln\frac{(M_W/m_n)(z)}{(M_W/m_n)(0)}+\frac{2}{11}\ln
\frac{(M_W/m_e)(z)}{(M_W/m_e)(0)}+\ln\frac{(M_X/\bar{M)}(z)}{(M_X/\bar{M})(0)}.\nonumber
\end{eqnarray}
We will make here the simplifying assumptions that the effects due to the variation
of $M_W/m_n$ and $M_W/m_e$ are subleading, as suggested by the more complete investigations
in \cite{CWTV} (in the linear approximation).

In the approximation of time-independent $M_W/m_n$ and $M_W/m_e$ the dependence of the fine
structure constant on redshift can be related directly to the $\beta$-function for the
grand unified gauge coupling (\ref{4}). For
$\Delta\alpha_{em}(z)=\alpha_{em}(z)-\alpha_{em}(0)$ one finds $\big(x=-\ln(1+z)\big)$
\begin{equation}\label{19}
\frac{\Delta\alpha_{em}(z)}{\alpha_{em}}=\frac{22\alpha_{em}}{7\alpha^2_X}\Delta\alpha_X(x)
\end{equation}
where $\alpha_X(x)$ obtains as a solution of
\begin{equation}\label{20}
\frac{\partial\alpha_X}{\partial x}=\beta_{\alpha}\frac{\partial\ln\chi}{\partial x}
=\frac{\beta_{\alpha}(\alpha_X,\delta)}{\sqrt{\delta}}
\sqrt{3\Omega_h(1+w_h)}.
\end{equation}
One observes that the $z$-dependence of $\Delta\alpha_{em}$ is particularly weak in the
region of large $\delta$ and $w_h$ close to $-1$. This will be our explanation why
$R=\Delta\alpha_{em}(z=0.13)/\Delta\alpha_{em}(z=2)$ is considerably smaller than expected
from a simple extrapolation linear in $z$. Crossover quintessence is precisely characterized
by large $\delta$ and $w_h\approx-1$ in the region of small $z$, whereas $\sqrt{1+w_h}
/\sqrt{\delta}$ is considerably larger for intermediate $z\approx 1.5-3$. One also observes
that for constant $(\Omega_h(1+w_h)/\delta)^{1/2}$ and $\eta=\beta_{\alpha}/\alpha$ the
dependence of $\Delta\alpha_{em}(z)$ is logarithmic in $1+z$. This is crucial for large
$z$ as for nucleosynthesis, where $\Delta\alpha_{em}$ turns always out to be much smaller
than expected from a linear extrapolation in redshift or time.

A quantitative estimate needs information about $\beta_{\alpha}$ - the form of the latter
may induce further nonlinearities. From our association of the scale $m$ with the scale
where $\alpha_X$ grows large the $\beta$-function (\ref{4}) would lead to a decrease of
$\alpha_X$ for decreasing $z$ if the last term $\sim S(\delta)$ is neglected. This would
result in a positive $\Delta\alpha_{em}(z=2)$, in contrast to the negative QSO value
(\ref{1}). Starting with large $\alpha_X(\chi=m)$ (we take $\alpha_X(\chi=m)=100$ for
definiteness) one would actually find values of $|\Delta\alpha_{em}(z=2)|$ much smaller
than the QSO value if $b_2$ exceeds $0.1$. This demonstrates that a fixed point behavior
for the running of $\alpha_X$ (and similar for other couplings) provides a powerful
explanation why the time variation of ``fundamental constants'' is actually a small effect!

There is, however, no reason why the fixed point value $\alpha_*$ should be precisley
independent of $\delta$. We parameterize a possible $\delta$-dependence by adding the last
term in eq. (\ref{4}) with
\begin{equation}\label{21}
S(\delta)=-\frac{\delta}{1+J_{\alpha}\delta}+6.
\end{equation}
This leads \footnote{We choose $S(\delta=0)>0$ such that for $b_6>0$ no new fixed
point for $\alpha_X$ is produced in the region of large $\alpha_X$. The value
$S(\delta=0)=6$ is unimportant.} to a crossover of the effective fixed point
$\alpha_*(\delta)$ between two values for small and large $\delta$. For
$b_6>0~,~J_{\alpha}>0$ the value of $\alpha_*(\delta)$ increases with
increasing $\delta$ and this will induce
$\Delta\alpha_{em}(z=2)<0$. At this stage the form of $\beta_{\alpha}$ is parameterized
by four parameters $b_2,b_4,B_6,J_{\alpha}$. Since the influence of the term
$S(\delta)$ on the location of the fixed point is small the ratio $b_2/b_4$ is
fixed by eq. (\ref{5}).

\begin{center}
{\bf Time variation of the fine structure constant}
\end{center}

At this point our model is fixed and we can now compute the detailed cosmological evolution
of the fine structure constant. The overall normalization of the time variation strongly
depends on $b_6$. As a benchmark we use eq. (\ref{1}). Therefore,
for given $b_2$ and $J_{\alpha}$ we fix $b_6$ such that
$\Delta\alpha_{em}(z=2)=-7\cdot10^{-6}$. The shape of $\Delta\alpha_{em}(z)$ depends
\footnote{In our case an increase of $|\beta_{\alpha}|$ for small $z$ partially cancels
the decrease of $(\Omega_h(1+w_h)/\delta)^{1/2}$.} now on $b_2$ and $J_{\alpha}$ as well
as on on $E$. In table 2 we show the values of $\Delta\alpha_{em}$ at $z=0.13,0.45,1100$ and
$10^{10}$ for several values of $(b_2,J_{\alpha},E)$.

\vspace{0.5cm}
\noindent
\label{table2a}
\begin{tabular}{|c|c|c|c|c|c|c|}\hline
model&
$z=0.13$&
$z=0.45$&
$z=1100$&
$z=10^{10}$&
$R$&
$\eta$ \\ \hline
A&
$-6\cdot 10^{-8}$&
$-2.6\cdot 10^{-7}$&
$-6.5\cdot 10^{-5}$&
$-7.8\cdot 10^{-5}$&
$0.0085$&
$5.1\cdot 10^{-14}$ \\ \hline
B&
$-6.2\cdot10^{-8}$&
$-2.7\cdot10^{-7}$&
$1.3\cdot 10^{-4}$&
$3.1\cdot 10^{-3}$&
$0.0088$&
$4.9\cdot 10^{-14}$ \\ \hline
C&
$-7.8\cdot10^{-8}$&
$-3.5\cdot 10^{-7}$&
$-2.5\cdot 10^{-5}$&
$-2.8\cdot 10^{-5}$&
$0.011$&
$8.9\cdot 10^{-14}$ \\ \hline
\end{tabular}

\vspace{0.5cm}
\noindent
Table 2a: Time variation of the fine structure constant $\Delta\alpha_{em}(z)/\alpha_{em}$
for various redshifts $z$, for $E=12$. The models for
$\beta_{\alpha}$ are $(A): J_{\alpha}=6,b_2=0.2,b_6=
0.84,~(B):J_{\alpha}=6,b_2=0.05,b_6=1.12,(C):J_{\alpha}=1,b_2=0.2,b_6=0.15$. For all models
$\Delta\alpha_{em}(z=2)/\alpha_{em}=-7\cdot 10^{-6}$. We also show the differential
acceleration $\eta$.

\vspace{0.5cm}
\noindent
\label{table2b}
\begin{tabular}{|c|c|c|c|c|c|c|}\hline
model&
$z=0.13$&
$z=0.45$&
$z=1100$&
$z=10^{10}$&
$R$&
$\eta$ \\ \hline
A&
$-1.4\cdot 10^{-7}$&
$-7.1\cdot 10^{-7}$&
$-4.4\cdot 10^{-5}$&
$-5.7\cdot 10^{-5}$&
$0.020$&
$4.3\cdot 10^{-14}$ \\ \hline
B&
$-2\cdot 10^{-7}$&
$-9.3\cdot10^{-7}$&
$1.7\cdot 10^{-4}$&
$3.6\cdot 10^{-3}$&
$0.028$&
$6.2\cdot 10^{-14}$ \\ \hline
C&
$-2\cdot 10^{-7}$&
$-9.9\cdot 10^{-7}$&
$-2.1\cdot 10^{-5}$&
$-2.3\cdot 10^{-5}$&
$0.029$&
$8.7\cdot 10^{-14}$ \\ \hline
\end{tabular}

\vspace{0.5cm}
\noindent
Table 2b: Same as table 2a, for $E=5$. The models for $\beta_{\alpha}$ are (A):
$J_{\alpha}=6,b_2=0.2,b_6=0.385,$ (B): $J_{\alpha}=6,b_2=0.05,b_6=0.783$,
(C): $J_{\alpha}=1,b_2=0.2,b_6=0.068$.

\vspace{0.5cm}

Consistency with the Oklo bound (\ref{2}) and the Re-decay bound (\ref{2a}) is found
for large $E$, in particular for large $J_{\alpha}$. It becomes clear that a
precise knowledge of $\Delta\alpha_{em}(z)$ could be used as a probe for
the dynamics of quintessence. For example, it seems very difficult to achieve
$R=|\Delta\alpha_{m}(z=0.13)/\Delta\alpha_{em}(z=2)|\leq 1/70$ for small values of
$E$, quite independently of the precise form of $\beta_{\alpha}$. The present investigation
confirms that a small value of $R$ requires $(1+w^{(0)}_h)\ll 1$. This property of the
equation of state seems to be favored by the cosmological tests (cf. table 1). We conclude
that the bound (\ref{AA2})
can be justified by our nonlinear analysis for CQ. The generic result that $w^{(0)}_h$
should be close to $-1$ does not depend on the precise values of $b_2$ and
$J_{\alpha}$. Increasing $b_2$ beyond $b_2=0.2$ (say $b_2=1$) induces almost no change.
On the other hand, for $b_2$ substantially smaller than $0.05$ we come to a region where
the fixed point $\alpha_{X,*}$ has not yet been reached with high precision and
$\Delta\alpha_{em}(z=2)$ is positive. For $b_2=0.05$ we have the interesting effect that
$\Delta\alpha_{em}(z)$ reaches a minimum near $z=2$, changing sign at somewhat higher
redshift. Only for the large value $E=12$ the functional dependence of $\Delta\alpha_{em}(z)$
deviates substantially from a linear behavior in $z$ in the range $0.5<z<2$, as seems to be
required by observation. For the models (A) (B) (cf. table 2) one finds a strong ``jump''
of $\Delta\alpha_{em}$ by a factor of $5$ between $z=1$ and $z=2$.

Let us next turn to the high values $z\approx 10^{10}$ characteristic for nucleosynthesis
and ask if the values for $\Delta\alpha_{em}$ quoted in table 2 are consistent with
observation. For nucleosynthesis, a change in $m_n/\bar{M}$ modifies the clock, i.e.
the rate of decrease of temperature in time units $\sim m^{-1}_n$. Furthermore, a change
in $\alpha_{em}$ affects the proton to neutron mass difference. Let us assume that the
time-variation of $m_n/\bar{M}$ is dominated by $\beta_{\alpha}$ via the change of the
strong gauge coupling (and therefore the ratio $\Lambda_{QCD}/M_X$). Including only the
first term in eq. (\ref{18}) yields \footnote{This relation differs from \cite{GUT}
where $M_W/M_X$ is kept fixed instead of fixed $m_n/M_W$ in the present work. The
quantitative difference is not very important, however.}
\begin{equation}\label{N1}
\frac{\Delta(m_n/\bar{M})}{(m_n/\bar{M})}=\frac{\pi}{11\alpha_{em}}
\frac{\Delta\alpha_{em}}{\alpha_{em}}=39.1\frac{\Delta\alpha_{em}}{\alpha_{em}}.
\end{equation}
A decrease of $m_n$ relative to $\bar{M}$ is equivalent to an increase of $\bar{M}$
at fixed $m_n$ (and thereby to a decrease of Newton's constant). In turn, this decreases
the Hubble parameter for a fixed temperature $T$ (measured in units of $m_n$). The net
effect is equivalent to the substraction of a (fractional) number of neutrino species.
We can thereby turn the limits on the effective number of neutrino species $\bar{N}_{\nu}$
into a limit on $\Delta\alpha_{em}/\alpha_{em}$ at $z\approx 10^{10}$. Assuming that the
primordial helium abundance $Y_p$ is in agreement with the value for time independent
couplings within an uncertainty $|\Delta Y_p/Y_p|<8\cdot 10^{-3}$ and using
$\Delta\ln Y_p=\frac{1}{3}\Delta\ln(m_n/\bar{M})$, yields the bounds
\begin{eqnarray} \label{N2}
|\Delta\ln(m_n/\bar{M})|&<&0.025
\end{eqnarray}
and
\begin{eqnarray} \label{N3}
|\frac{\Delta\alpha_{em}(z=10^{10})}{\alpha_{em}}|&<&6.4\cdot 10^{-4}.
\end{eqnarray}
We emphasize that an effective neutrino number $\bar{N}_{\nu}<3$ would be a clear signal
for a time variation of couplings! We also observe that bounds on $m_n/\bar{M}$ as
(\ref{N2}) can be interpreted equivalently as bounds on the time variation of Newton's
constant (with $m_n$ fixed).

We may compare the effect of the change of $m_n/\bar{M}$ with the effect of the change in
the proton-neutron mass difference $\delta m=m_n-m_p$ which contributes $\Delta\ln Y_p
\approx\frac{1}{3}\Delta\ln(m_n/\bar{M})-\Delta \ln\delta m$. Keeping the ratios of up-and
down quark masses to $m_n$ fixed one has \cite{Gas} $\Delta\ln\delta m\approx-0.6
\Delta\ln\alpha_{em}$. The contribution to $\Delta\ln Y_p$ is far less than the contribution
from $\Delta\ln(m_n/\bar{M})$. This explains why the bound (\ref{N3}) is more restrictive
than several previous bounds \cite{BBN}, \cite{Av} concentrating on the effect from
$\Delta\ln\delta m$. We are aware that the variation of $\Delta\ln(m_n/\bar{M})$ may be
affected by the neglected terms in eq. (\ref{18}). If there are no particular cancellations,
however, the bound (\ref{N3}) should remain of the same order of magnitude.

The size of $\Delta\alpha_{em}$ at nucleosynthesis and even the sign depend critically
on $b_2$ (cf. table 2). If we adopt the bound
\begin{equation}\label{N4}
|\frac{\Delta\alpha_{em}(z=10^{10})}{\alpha_{em}}|<10^{-3}
\end{equation}
we infer that $b_2$ should be larger than $0.05$. It is striking that the values of
$\Delta\alpha_{em}/\alpha_{em}$ at nucleosynthesis are very far from an extrapolation
with $\partial\ln\alpha_{em}/\partial t=const.$ or even from $\partial\ln\alpha_{em}/
\partial\ln a=const.$. The nonlinearity in $\partial\ln\alpha_{em}/\partial\ln a$ is
mainly due to the attraction of $\alpha_X$ to a $\delta$-dependent fixed point. Both for
$E=5,12$ the models with sufficiently large $b_2$ meet the bounds (\ref{N3}) or
(\ref{N4}) (model A). They are also consistent with bounds from the CMB unisotropies
\cite{Av}.

Precision experiments with atomic clocks measure the variation $\Delta\alpha_{em}/\alpha_{em}$
per year
\begin{eqnarray}\label{Z1}
\frac{\dot{\alpha}_{em}}{\alpha_{em}}[yr^{-1}]&=&-H[yr^{-1}]
\frac{\partial\ln\alpha_{em}}{\partial\ln(1+z)}\nonumber\\
&=&-5.44\cdot 10^{-10}\frac{\Delta\alpha_{em}(z=0.13)}{\alpha_{em}}yr^{-1}.
\end{eqnarray}
Here the Hubble parameter is expressed in units of $yr^{-1}$ and we use the observation
that for our models $\partial\ln\alpha_{em}/\partial x$ is essentially constant in the range
$0<z<0.13$. Measuring the values quoted in table 2 would need a considerable improvement of
the present accuracy \cite{Sor} $\dot{\alpha}_{em}/\alpha_{em}=(4.2\pm6.9)\cdot 10^{-15}yr^{-1}$.
Using eq. (\ref{N1}) we obtain a similar formula for the variation of Newton's constant in
the range of low redshift,
\begin{equation}\label{ZZ}
\frac{\dot{G}}{G}=78.2\frac{\dot{\alpha}_{em}}{\alpha_{em}}=-4.25\cdot 10^{-8}
\frac{\Delta\alpha_{em}(z=0.13)}{\alpha_{em}},
\end{equation}
safely below the present bounds.

In our approximation where the dominant field dependence of the various couplings
arises from $\beta_{\alpha}$ we can also estimate the size of the differential acceleration
$\eta$ between two test bodies with equal mass but different composition. One finds
\cite{CWTV}
\begin{equation}\label{22}
\eta=5\cdot10^{-2}\left(\frac{\beta_{\alpha}}{\alpha_X}\right)^2
\delta^{-1}\Delta R_Z
\end{equation}
where for typical experimental tests of the equivalence principle $\Delta R_Z=\Delta Z/
(Z+N)\approx 0.1$. Here $\beta_{\alpha}/\alpha_X$ and $\delta$ have to be evaluated at
$z=0$. Results for $\eta$ are also displayed in table 2. We conclude that our CQ-model
is compatible with the present experimental bounds \cite{EP}
$|\eta|<3\cdot 10^{-13}$ for most choices of
$\beta_{\alpha}$. Nevertheless, the models with $J_{\alpha}=1$ show that already now the
tests of the equivalence principle can help to discriminate between different models!
It is obvious that further improvements of the accuracy of tests
of the equivalence principle should either confirm or reject the interpretation of the
QSO-results within our CQ-model.

In summary, we have found models of crossover quintessence that can explain the QSO value
for the time dependence of the fine structure constant and are nevertheless compatible
with all observational bounds on the time variation of couplings and tests of the
equivalence principle. This explanation does not involve any particular cancellation of
effects from the variation of different couplings. The parameter set $E=12,b_2=0.2,
J_{\alpha}=6$ may serve as an illustration. The same models do also very well with all
present cosmological tests!

\begin{center}
{\bf Variation of electron-proton mass ratio}
\end{center}

Let us finally ask if our assumption that $\Delta\alpha_{em}$ is dominated by the change
of $\alpha_X$ can be justified. In addition to the arguments presented in \cite{CWTV}
we may also have a look at the recently reported result \cite{Iv} on the variation of
$m_e/m_p$ in the range $z=2.3-3$
\begin{equation}\label{X1}
\Delta\left(\frac{m_e}{m_p}\right)/\left(\frac{m_e}{m_p}\right)=(-5.7\pm3.8)\cdot 10^{-5}.
\end{equation}
The deviation from zero is statistically not very significant. Nevertheless, it is
interesting to know if such an effect would influence the time history of the fine structure
constant and the tests of the equivalence principle. We therefore
investigate what would be the effect on
$\Delta\alpha_{em}$ of a variation $\Delta\ln(m_e/m_n)(z=2)
=-6\cdot 10^{-5}$. Keeping $M_W/m_n$ fixed would yield a
contribution (cf. eq. (\ref{17}))
\begin{equation}\label{X2}
\frac{\Delta\alpha_{em}(z=2)}{\alpha_{em}}=\frac{2\alpha_{em}}{\pi}
\Delta\ln\left(\frac{m_e}{m_n}\right)\approx-2.8\cdot 10^{-7}.
\end{equation}
This amounts only to $4\%$ of the QSO value (\ref{1}). Similarly, the contribution
(cf. eq. (\ref{18})) to
\begin{equation}\label{X3}
\ln\frac{(m_n/\bar{M})(z=2)}{(m_n/\bar{M})(z=0)}\approx-\frac{2}{11}\Delta\ln
\left(\frac{m_e}{m_n}\right)\approx 1.1\cdot 10^{-5}
\end{equation}
is only $4\%$ of the effect due to $\Delta\alpha_{em}$, i.e.
\begin{equation}\label{X4}
\ln\frac{(m_n/\bar{M})(z=2)}{(m_n/\bar{M})(z=0)}=\frac{\pi}{11\alpha_{em}}
\frac{\Delta\alpha_{em}(z=2)}{\alpha_{em}}\approx-2.7\cdot 10^{-4}.
\end{equation}
Neglecting those contributions seems therefore well justified. Comparison of eqs.
(\ref{X4}) and (\ref{X1}) incidentally shows that the value (\ref{X1}) corresponds
roughly to the size of the effect that would be expected in absence of particular
cancellations - with $\ln(m_e/\bar{M})(z=2)/\ln(m_e/\bar{M})(z=0)
\approx-3.3\cdot 10^{-4}$. The effect of $\Delta\ln(m_e/m_n)$ on the tests of the
equivalence principle is more sizeable. Adding this effect to the differential acceleration
multiplies $\eta$ in eq. (\ref{22}) by a factor $1+\tilde{Q}$ \cite{CWTV}
\begin{equation}\label{X5}
\tilde{Q}\approx 10\frac{d(m_e/m_n)}{d\alpha_{em}}_{|z=0}\approx 10
\frac{\Delta(m_e/m_n)}{\Delta\alpha_{em}}(z=2)\approx 6~,
\end{equation}
where the quantitative estimate neglects a possible $z$-dependence of $\tilde{Q}$.
Multiplication of the $\eta$-values in table 2 by a factor 7 shows that for part of the
models a value $\Delta\ln(m_e/m_p)(z=2)=-6\cdot 10^{-5}$ enters in conflict with the tests
of the equivalence principle. The models with $E=12$ and $J_{\alpha}=6$ may be considered
as borderline in view of the possibility of partial cancellations with effects from the
variation of other mass ratios.

\newpage
\begin{center}
{\bf The future universe}
\end{center}

If we live today a crossover period of our universe, to which future will the transition
lead? In our model, the answer depends in important aspects on the behavior of
$\beta_{\delta}$ for large $\delta$. The scenario that we have discussed first is
characterized by ever increasing $\delta$ (or the approach to a fixed point
($1/\delta)_*=0)$.
In this case $\Omega_h$ will grow towards one and $w_h$ approaches $-1$ in future times.
In these respects the universe will resemble asymptotically a universe characterized by
a cosmological constant. Nevertheless, if $k(\phi)$ remains finite for all finite $\phi$,
the dark energy density $\rho_h$ will not approach a constant since $w_h+1$ will always
remain positive, approaching zero only for $\phi\rightarrow\infty$. An asymptotic behavior
$\beta_{\delta}=D\delta$ for large $\delta$ corresponds to $k(\phi)=k_C\exp
\left\{\frac{D}{4}(\phi-\phi_c)\right\}$. This is the asymptotic behavior of inverse power
law quintessence \cite{Rat}: rescaling the cosmon kinetic term leads \footnote{
For our ansatz (\ref{9}) with $D=E/J_{\delta}$ we find very small $\alpha$, i.e. for
$E=5,J_{\delta}=0.05$ follows $\alpha=0.04$.} to a potential
$V(\tilde{\varphi})\sim\tilde{\varphi}^{-\alpha}~,~\alpha=4/D$. In leading order one finds
$1+w_h=1/(3k^2)$. The potential energy decreases logarithmically with $a,~V\sim(\ln a)^{-2/D}$
and we infer the asymptotic behavior $\Omega_m\sim(\ln a)^{2/D}/a^3,\Omega_h=1-\Omega_m$.

As a second scenario we discuss the alternative that $\beta_{\delta}$ has a zero at some
finite value $\delta_*$. As a natural choice we take $\delta_*=6$, noting that for
$\delta=6$ the kinetic term for $\chi$ in eq. (\ref{3}) vanishes, and consider
\begin{equation}\label{V1}
\beta_{\delta}=E\delta^2\left(1-\frac{\delta}{6}\right).
\end{equation}
Our cosmological model describes now a crossover between the two fixed points $\delta_*=0$
and $\delta_*=6$, being relevant for the early and late universe, respectively. The
solution $\delta(\chi)$ to eq. (\ref{V1}) obeys
\begin{equation}\label{V2}
\frac{\delta}{1+\frac{\delta}{6}\ln\left(\frac{1}{\delta}-\frac{1}{6}\right)}=
\frac{1}{E\ln(\chi_c/\chi)}
\end{equation}
and approaches the solutions of eqs. (\ref{6}) or (\ref{9}) for small $\delta$. The
asymptotic cosmology for large $t$ is different, however. It corresponds to ``power law
inflation'' where the equation of state approaches the asymptotic value
$w^{\infty}_h=-7/9$, with
\begin{eqnarray}\label{V3}
V&=&\frac{24\bar{M}^2}{t^2}~,~T=\frac{\bar{M}^2}{2}\dot{\phi}^2=\frac{3\bar{M}^2}{t^2}~,
H=\frac{3}{t}\nonumber\\
\delta&=&6-\frac{36}{e}\exp[-3E(\phi-\phi_c)].
\end{eqnarray}
The Hubble parameter decreases $\sim t^{-1}$, with $a\sim t^3$. Comparing
$\rho_h\sim a^{-2/3},\rho_m\sim a^{-3}$ we conclude that $\Omega_h\rightarrow 1,\Omega_m
\sim a^{-7/3}$. As for the first scenario, the future universe will be completely
dominated by dark energy. However, the dynamics of the dominant quintessence component
differs between the two scenarios.

We have displayed the characteristic cosmological observables in table 3.
In order to demonstrate the spread in this type of models we have chosen
$h=0.65$ and $\Omega^{(0)}_h=0.7$ and also display smaller values of $E$ as compared to
table 1.

\vspace{0.5cm}
\noindent
\label{table3}
\begin{tabular}{|c|c|c|c|c|c|}\hline
E&
$w^{(0)}_h$&
$t^{(0)}/10^{10}yr$&
$\Omega^{(ls)}_h$&
$l_3$&
$\sigma_8/\sigma^{(\Lambda)}_8$ \\ \hline
$2$&
$-0.65$&
$1.31$&
$0.045$&
$768$&
$0.46$ \\ \hline
$5$&
$-0.81$&
$1.37$&
$0.019$&
$795$&
$0.69$ \\ \hline
\end{tabular}

\vspace{0.5cm}
\noindent
Table 3: Characteristic cosmological quantities for CQ-models with fixed point at
$\delta_*=6$.

We also show in  table 4 the time variation of $\alpha_{em}$,
using in eqs. (\ref{4})(\ref{21})
$J_{\alpha}=0$ and tuning $b_6$ such that $\Delta\alpha_{em}(z=2)/\alpha_{em}=-7\cdot
10^{-6}$.

\vspace{0.5cm}
\noindent
\label{table4}
\begin{tabular}{|c|c|c|c|}\hline
E&
$b_2$&
$b_6$&
$\Delta\alpha_{em}(z=0.13)$\\ \hline
$2$&
$0.2$&
$0.011$&
$-1.1\cdot10^{-6}$ \\ \hline
$5$&
$0.06$&
$0.022$&
$-1.8\cdot 10^{-6}$ \\ \hline
$5$&
$0.2$&
$0.01$&
$-9.5\cdot 10^{-7}$ \\ \hline
\end{tabular}

\vspace{0.5cm}
\noindent
Table 4: Variation of $\Delta\alpha_{em}(z=0.13)/\alpha_{em}$ for fixed $\Delta\alpha_{em}
(z=2)/\alpha_{em}=-7\cdot 10^{-6}$. The models for $\beta_{\alpha}$ use $J_{\alpha}=0$.

\vspace{0.5cm}
\noindent
It becomes obvious that the bound (\ref{2}) is not obeyed. Consistency with the Oklo-natural
reactor results would therefore require a substantial cancellation between effects from
the time variation of different couplings.

\vspace{0.5cm}
\begin{center}
{\bf Testing the model}
\end{center}

Within a rather wide class of quintessence models we have seen how a combination of the
QSO-result (\ref{1}) with the bounds on $\dot{\alpha}_{em}$ at lower redshift (\ref{2})
(\ref{2a}) restricts the allowed parameter space. The present status of observations favors
a rather sharp crossover for quintessence, where the equation of state has shifted from a
moderately negative value at $z\approx 2$ to a value very close to $-1$ for the recent
cosmological epoch. An example is given by the effective action (\ref{3}) with running
couplings determined by the $\beta$-functions (\ref{9})(\ref{4})(\ref{21}), and parameters
$E=12,J_{\delta}=0.05,b_2=0.2,b_4=8,b_6=0.84,J_{\alpha}=6$ (first line in table 1, model
A in table 2a). Without the need of excessive tuning of parameters or invoking cancellation
effects from the time variation of different couplings, this demonstrates the existence of a
field theoretical model that is consistent with all data on the cosmological evolution and the
time variation of couplings.

What will be the tests of this model? First of all, the QSO result should be confirmed by
new independent data. Hopefully this will allow for more precise restrictions on the shape of
the function $\Delta\ln\alpha_{em}(z)$ which can be directly compared with the prediction of
this model. Second, an improvement of the accuracy of the tests of the equivalence principle
by an order of magnitude should lead to a direct detection of the new interaction mediated
by the cosmon field. In view of its important role in present cosmology we may call this
fifth force (besides gravity electromagnetism, weak and strong interactions) ``cosmo-dynamics''.
Third, cosmological tests should distinguish crossover quintessence from a cosmological
constant. The main signature in this respect is probably not the equation of state of dark
energy in the most recent epoch (say for $0<z<0.5)$. The value of $w^{(0)}_h$ in table 1
shows that this quantity may be quite near to the value for a cosmological constant (i.e.
$w^{(0)}_h=-1$). If the QSO-results on the time variation of the fine structure constant are
to be explained by quintessence this requires a substantial evolution of the cosmon field in
the epoch near $z=2$ (as compared to the rate of evolution today). We therefore expect that
$w_h(z)$ increases with increasing $z$, leading to $\Omega_h(z)$ substantially larger than
for a cosmological constant during structure formation and last scattering. Test of structure
formation and precision tests on the CMB may be able to measure the value of $\sigma_8/
\sigma^{(\Lambda)}_8$ and $\Omega^{(ls)}_h$ in table 1. Finally, an improvement of the
accuracy of atomic clocks by three orders of magnitude would replace the bounds
(\ref{2})(\ref{2a}) within an experimentally well controlled setting.

\end{document}